\RequirePackage[patch]{kvoptions} 

\documentclass{mtits_paper}

\usepackage{units} 

\providecommand{\bc}{\begin{center}}
\providecommand{\ec}{\end{center}}
\providecommand{\be}{\begin{equation}}
\providecommand{\ee}{\end{equation}}
\providecommand{\bea}{\begin{eqnarray}}
\providecommand{\eea}{\end{eqnarray}}
\providecommand{\bdm}{\begin{displaymath}}
\providecommand{\edm}{\end{displaymath}}
\providecommand{\bdma}{\begin{eqnarray*}}
\providecommand{\edma}{\end{eqnarray*}}
\providecommand{\ba}{\begin{eqnarray*}}
\providecommand{\ea}{\end{eqnarray*}}
\providecommand{\bi}{\begin{itemize}}
\providecommand{\ei}{\end{itemize}}
\providecommand{\benum}{\begin{enumerate}}
\providecommand{\eenum}{\end{enumerate}}

\providecommand{\refkl}[1]{(\ref{#1})}

\providecommand{\twoCases}[4]{
  \left\{ 
    \begin{array}{ll} 
      #1 & #2 \\
      #3 & #4 
    \end{array} 
  \right.
}

\providecommand{\text}[1]{{\mbox{ #1}}}

\providecommand{\ablpart}[2]{\frac{\partial #1}{\partial #2}}  
\providecommand{\abl}[2]{\frac{{\rm d} #1}{{\rm d} #2}}  

\providecommand{\sub}[1]{_{\rm #1}}
\renewcommand{\sup}[1]{^{\rm #1}}

\begin{document}


\begin{abstract}
Vehicle-infrastructure communication opens up new ways to improve
traffic flow efficiency at signalized 
intersections. In this study, we assume that equipped vehicles can
obtain information about switching times of relevant traffic lights in
advance, and additionally counting data from upstream detectors. By means of simulation, we
investigate, how equipped
vehicles can make use of this information to improve traffic flow.
Criteria include cycle-averaged capacity, driving comfort, fuel
consumption, travel time, and the number of stops. Depending on the
overall traffic demand and the penetration rate of equipped vehicles,
we generally find several percent of improvement.
\end{abstract}

\section{Introduction}
%
Individual vehicle-to-vehicle  and vehicle-to-infrastructure
communication, commonly refer\-red to as V2X,  are
novel components of intelligent-traffic systems
(ITS). Besides more traditional ITS applications
such as variable speed limits on freeways or
traffic-dependent signalization~\cite{SCOOT,SCATS}, V2X promises new
applications to make traffic flow more efficient or driving more
comfortable and economic.
While there are many investigations
focussing on technical issues such as connectivity given a certain hop
strategy, communication
range, and percentage of equipped
vehicles (penetration rate), e.g., \cite{thiemann-IVC-PRE08}, few papers
have investigated actual strategies to improve traffic flow characteristics. On
freeways, a jam-warning system based on communications to and from
road-side units (RSUs) has been proposed~\cite{Kranke-Fisita2008}. Furthermore, a
traffic-efficient adaptive-cruise control (ACC) has been proposed
which relies on V2X communcation to determine the local traffic
context influencing, in turn, the ACC
parameterization~\cite{kesting-acc-roysoc}. Regarding city traffic,
early forms of V2X have been investigated in the European projects
Prometheous/Drive~\cite{PrometheusDrive1991}. However, these
initiatives were more
focussed on safety and routing information without explicitely
treating any interactions with traffic lights. The investigation which is
arguably most related to our work is the
thesis~\cite{Otto2011kooperative} on cooperative traffic control in
cities discussing in depth the possibly destructive
interplay between V2X (traffic-dependent signalization) and X2V
(driver information relying on predetermined signalization).

In this contribution, we focus on city traffic at signalized
intersections and investigate a set of strategies that is complementary to
the self-controlled signal control strategy of L\"ammer and Helbing~\cite{lammer2008SelfControl}: While, in the
latter, the traffic lights ``know'' the future traffic, we assume that
equipped vehicles know the future states of the next traffic
light. In principle, the resulting traffic-light assistant (TLA) can
operate in the information-based manual mode, or in the ACC-based
automatic mode on which we will focus in this work.

In the next section, we lay out the methodology of this
simulation-based study and define the objectives. Section~3
presents and analyzes the actual strategies ``economic approach'',
``anticipative start'', and ``flying start''.
In the concluding Section~4, we discuss the results and point at
conditions for implementing the strategies in an actual TLA.

\section{Methodology}

\subsection{Car-Following Model}
%
In order to get valid results, the underlying car-following model must
be (i) sufficiently realistic to represent ACC driving in the
automatic mode of the TLA, (ii) simple enough for
calibration, and (iii) intuitive enough to readily implement the new
strategies by re-parameterizing or augmenting the model. We apply the
``Improved Intelligent-Driver Model''
(IIDM) as described in  Chapter~11 of the
book~\cite{TreiberKesting-Book}. As the original Intelligent-Driver
Model (IDM)~\cite{Opus}, it is a
time-continuous car-following model with a smooth acceleration
characteristics. Assuming speeds $v$ not exceeding the desired speed
$v_0$, its acceleration equation as a
function of the (bumper-to-bumper) gap $s$, the own speed $v$ and the speed $v_l$ of the
leader reads
\begin{equation}
\label{IIDM}
\abl{v}{t}:=a\sub{IIDM}(s,v,v_l) 
=\twoCases{(1-z^2)\, a}
{\quad z=\frac{s^*(v,v_l)}{s}\ge 1,}
{ \left(1-z^{\frac{2a}{a\sub{free}}}\right)\, a\sub{free}}
{\quad \text{otherwise,}}
\end{equation}
where the expressions for the desired dynamic gap $s^*(v,v_l)$ and the
free-flow acceleration $a\sub{free}(v)$ are the same as that of the
IDM,
\bea
\label{sstar}
s^*(v,v_l) &=& s_0
+\max\left[vT+\frac{v(v-v_l)}{2\sqrt{ab}}, \, 0\right],\\
a\sub{free}(v) &=& a
\left[1-\left(\frac{v}{v_0}\right)^{\delta}\right].
\eea
The IIDM has the same parameter
set as the IDM:
desired speed $v_0$, 
desired time gap $T$, minimum space gap $s_0$, desired
acceleration $a$, and desired deceleration $b$.  However, it resolves
two issues of the basic IDM when using it as an ACC acceleration
controller: (i) the IIDM time-gap parameter $T$ describes exactly the time
gap in steady-state car-following situations while the actual IDM
steady-state time gaps are somewhat larger~\cite{Opus}, (ii) a platoon of
vehicle-drivers with same desired speed $v_0$ will not disperse over time
as would be the case for the IDM. Notice that a slightly different
formulation, the ``IDM plus'' with the acceleration function
$a\sub{IDM+}(s,v,v_l)=\min[a\sub{free}, (1-z^2)a]$, would serve this
purpose as well.
 
By describing the vehicle motion with a time-continuous car-following
model, we have neglected the in-vehicle
control path since such models implicitely
reflect an acceleration response time of zero. 
It might be necessary to explicitely model vehicle
responses by a sub-microscopic model (e.g., PELOPS) when actually
deploying such a system.

\begin{figure}[!ht]
  \begin{center}
    \includegraphics[width=0.7\textwidth]{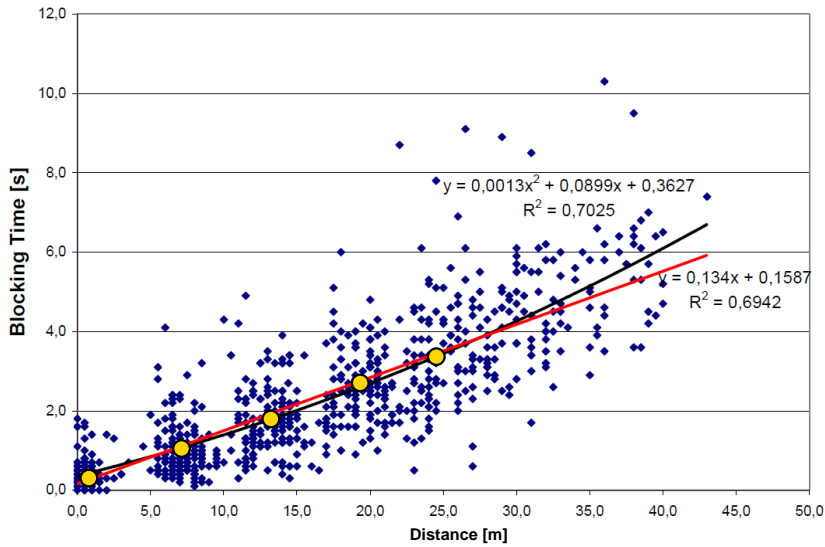}
  \end{center}
  \caption{\label{fig:calibr}Calibration of the microscopic model with respect to the
    starting times and positions of a queue of waiting vehicles
    relative to the begin of the green phase (solid circles). Data are of the
    measurements in~\cite{Kuecking}.}
\end{figure}

\subsection{\label{sec:calibr}Calibration}
%

Since we will
investigate platoons travelling from traffic light to traffic light,
the acceleration model parameters $v_0$, $T$, $s_0$, $a$, and $b$ and
the vehicle length $l\sub{veh}$ (including their variances) should be
calibrated to data of starting and stopping situations.

For calibrating $a$, $T$, and
the combination $l\sub{eff}=l\sub{veh}+s_0$ (effective vehicle
length), we use the empirical
results of K\"ucking~\cite{Kuecking} taken at three intersections in
the city of Hannover, Germany. There,
the ``blocking time'' of the $n\sup{th}$ vehicle of a waiting
queue (the time interval this vehicle remains stopped after the light
has turned green) has been measured vs. the distance of this vehicle to the stopping
line of the traffic light. Figure~\ref{fig:calibr} reproduces these
data together with the simulation results (orange bullets) for the
calibrated parameters $l\sub{eff}=\unit[6.5]{m}$,
$a=\unit[1.5]{m/s^2}$, and $T=\unit[1.2]{s}$ assuming 
identical driver-vehicles. Further simulations with heterogeneous
drivers and vehicles reveal that independently and uniformly distributed values for
$l\sub{eff}$, $T$ and $a$ with standard deviations of the order of
\unit[30]{\%} of the respective expectation value can reproduce the
observed data scatter and its increase with the vehicle position (for
positions $n=5$ and higher, the scattering does no longer allow to
identify $n$). Moreover, since trucks are excluded from the
measurements, it is reasonable to assume that the observed cars have
an average length of \unit[4.5]{m} resulting in an expectation value
$s_0=\unit[2]{m}$ for $s_0$.

For estimating  the comfortable deceleration, the approach to a red
traffic light is relevant. Investigations on the Lankershim data set
of the NGSIM data~\cite{NGSIM} including such situations show that a
typical deceleration is
$b=\unit[2]{m/s^2}$~\cite{viti2010microscopic}.
Finally, for the desired speed, we assumed a fixed value of
$v_0=\unit[50]{km/h}$ representing the usual inner-city speed limit in Germany.

\subsection{\label{sec:sim}Simulation}

\begin{figure}[!ht]
  \begin{center}
    \includegraphics[width=0.6\textwidth]{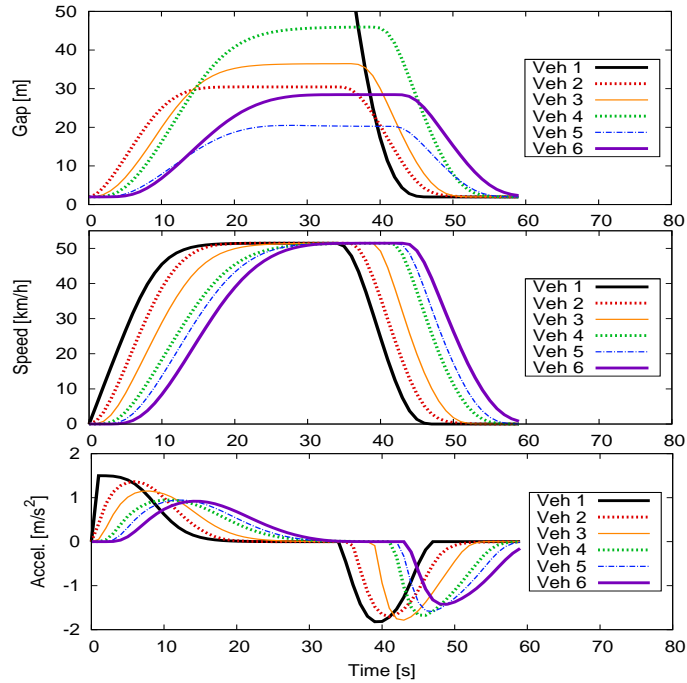}
  \end{center}
  \caption{Start and stop of the simulated platoon of heterogeneous
    vehicle-drivers in the reference case.
  }
  \label{fig:reference}
\end{figure}

While the parameters clearly are distributed due to inter-vehicle and inter-driver
variations, it is nevertheless necessary to use the same vehicle
population for all the following simulation experiments. Specifically,
we use following sequence of four vehicle-driver combinations: 1. average driver
(expectation values for the parameters),
2. agile driver ($a$ increased to $\unit[2]{m/s^2}$, $T=\unit[1.8]{s}$),
3. less agile but anticipative driver ($a$ and $b$ decreased to
$\unit[1.2]{m/s^2}$ and $\unit[1]{m/s^2}$, respectively), 
and 4. a truck
($l\sub{veh}=\unit[12]{m}$, $T=\unit[1.7]{s}$, and
$a=b=\unit[1]{m/s^2}$).
 If necessary, this sequence is
  repeated. Figure~\ref{fig:reference} shows the simulation result for
  the start-and-stop reference scenario against which the strategies of
  the traffic light assistent will be tested in the next section.

\subsection{Traffic Flow Metrics}

In the ideal case, the TLA reduces the travel time of the equipped and
the other vehicles, increases driving comfort and traffic flow
efficiency, and reduces fuel
consumption. To assess travel
time, we use the average speed of a vehicle, or average over all vehicles during
the complete simulation run. As proxy for the driving comfort, we take
the number
of stops during one simulation, or, equivalently, the fraction of
stopped vehicles. Traffic flow efficiency is equivalent to the
cycle-averaged dynamic capacity, i.e., the average number of vehicles
passing a traffic light per cycle in congested congestions in the
absence of gridlocks. Finally,  we determine the fuel consumption
 by a physics-based modal
consumption model as described in Chapter~20.4 of the
book~\cite{TreiberKesting-Book}). Such models take the simulated
trajectories and some vehicle attributes 
as input and return the instantaneous consumption rate
and the total consumption of a given vehicle. To be specific, we
assume a mid-size car with following attributes: 
Characteristic map of a \unit[118]{kW} gasoline engine as in Fig.~20.4
of~\cite{TreiberKesting-Book}, idling
power $P_0=\unit[3]{kW}$, total mass $m=\unit[1\,500]{kg}$,
friction coefficient $\mu=0.015$, air-drag coefficient $c_d=0.32$,
frontal cross-section $A=\unit[2]{m^2}$, a dynamic tyre radius
$r\sub{dyn}=\unit[0.286]{m}$. Furthermore, we assume a five-gear transmission with
transmission ratios of 13.90, 7.80, 5.25, 3.79, and 3.09,
respectively, and chose
the most economic gear for a given driving mode characterized by $v$
and $\abl{v}{t}$. The engine power management includes overrun-fuel
cutoff, idling when the vehicle is stopped, and no energy recuperation
during braking.

\section{Strategies of the Traffic Light Assistant and their Simulation}
%
The appropriate TLA strategy depends essentially on the arrival time
at the next traffic light
relative to its phases.  We
distinguish following approaching situations
(cf. Fig.~\ref{fig:KOLINE}):

\begin{figure}[!ht]
  \begin{center}
    \includegraphics[width=0.7\textwidth]{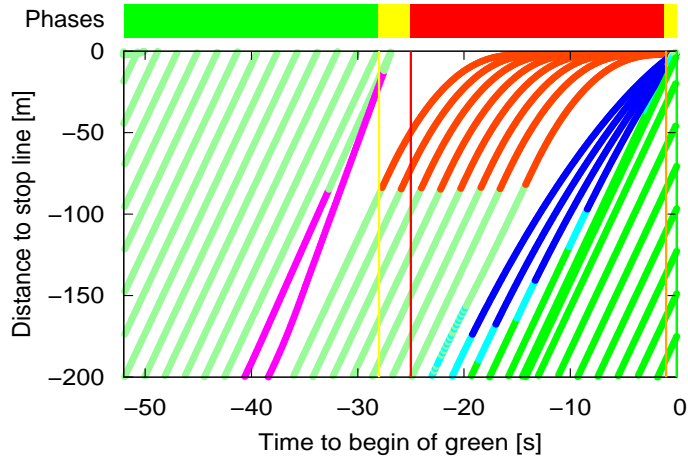}
  \end{center}
  \caption{Approach situations relative to the phases of
    the traffic light. Each trajectory
    corresponds to an individual simulation of the considered vehicle
    with no interactions to other vehicles. For the color coding, see
    the main text.}
  \label{fig:KOLINE}
\end{figure}

\bi
\item A stop is unavoidable (the first seven of the red trajectories
  of Fig.~\ref{fig:KOLINE}),\\[-2em]
\item anticipative start compensating for the reaction time of the
  first vehicle (last red trajectory),\\[-2em]
\item flying start realized by anticipative braking (blue
  trajectories),\\[-2em]
\item free passage (green trajectories),\\[-2em]
\item temporary ``boost'' to catch the last part of the green phase
  (violet trajectories).  
\ei
Nothing can be done in the situation of a free passage while the
``boost'' strategy implies temporarily exceeding the speed
limit. Therefore, we
will only develop and simulate strategies for the first three situations. Generalizing the above sketch, we
will  also investigate how other (equipped or
non-equipped) vehicles will affect the strategies. Furthermore, by a
complex simulation over several cycles, we investigate
any (positive or negative) interactions between the strategies and
between equipped and non-equipped vehicles.

\subsection{\label{sec:approachStop}Approach to a Stop}

In certain situations, a stop behind a red light or a waiting queue is
unavoidable. This situation is true if (i) extrapolated
constand-speed arrival occurs during a red phase, and (ii) the
``flying-start'' strategy of Sect.~\ref{sec:flyingStart} would produce
minimum speeds below a certain threshold which we assumed to be
$v\sup{flying}\sub{min}=\unit[10]{km/h}$. Notice that this szenario
may also apply for approaching green traffic lights if the car cannot make it
to the traffic light before switching time: In such a situation, drivers of non-equipped
cars would just go ahead braking later and
necessarily harder.
While this situation is not relevant for improving flow efficiency, it is
nevertheless possible to reduce fuel consumption by early use of the engine
brake, i.e., early activation of the overrun cut-off.

\begin{figure}[!ht]
  \begin{center}
    \includegraphics[width=0.95\textwidth]{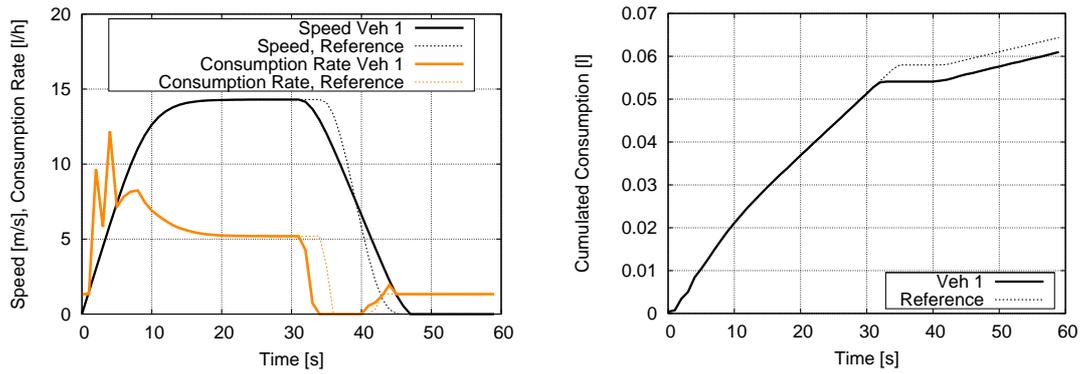}
  \end{center}
  \caption{\label{fig:approachStop}Fuel-saving approach to a waiting
    queue. Left: speed profile and 
    instantaneous consumption rate; right: cumulative consumption
    during the complete start-stop cycle.}
\end{figure}

In the car-following model, we implement this strategy by 
reducing the comfortable deceleration from $b=\unit[2]{m/s^2}$ to
$\unit[1]{m/s^2}$ (homogeneous driver-vehicle population), or by
\unit[50]{\%} for each vehicle (heterogeneous population). Reducing
the desired deceleration means earlier braking, in line with this strategy.

Figure~\ref{fig:approachStop} shows speed and consumption profiles for
an equipped vehicle (solid lines) vs. the reference (dotted). The equipped vehicle
itself saves about \unit[3.5]{ml} of fuel (\unit[6]{\%} for the
complete start-stop cycle). The two next (non-equipped)
followers save about
\unit[3]{\%} and \unit[1]{\%}, respectively.

\subsection{\label{sec:anticipativeStart}Anticipative Start}

\begin{figure}[!ht]
  \begin{center}
    \includegraphics[width=0.9\textwidth]{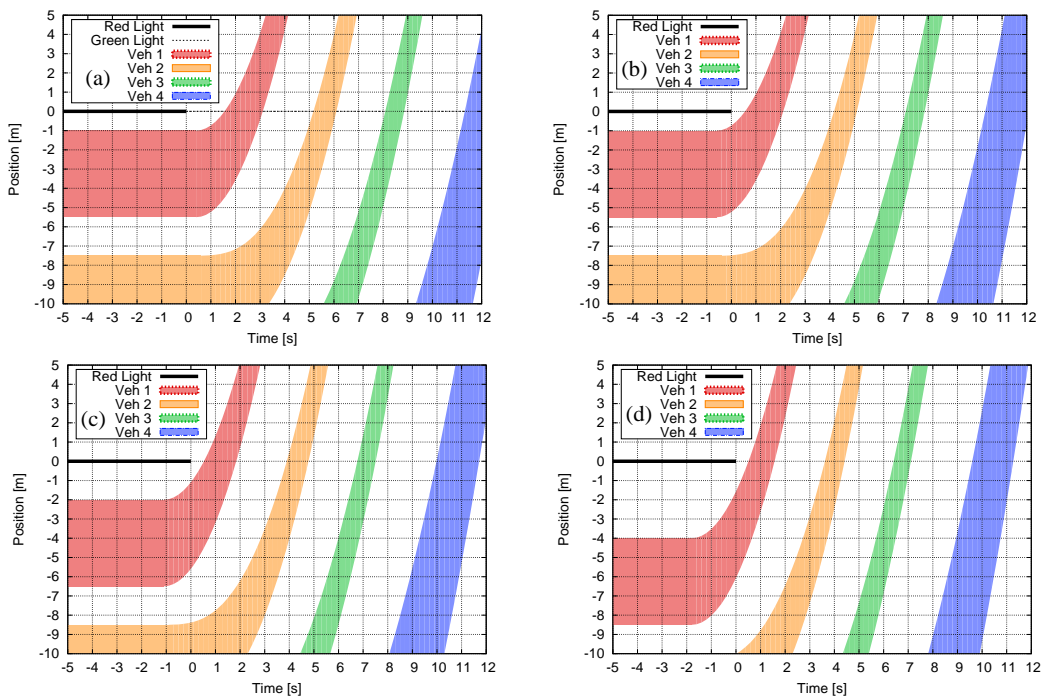}
  \end{center}
  \caption{\label{fig:stopStart}Start at green from the first position of a queue of
    waiting vehicles. (a) reference; (b) anticipative
    start; (c) anticipative start plus \unit[1]{m} additional
    gap; (d) anticipative start plus \unit[3]{m} additional
    gap.}
\end{figure}

The rationale of the strategy of the anticipative start is to
compensate for the reaction time delay $\tau$.
Since the reaction time is only relevant
for the driver of the first vehicle in a queue, the anticipative-start
strategy is
restricted to this vehicle. In the reference case corresponding
to the calibrated parameters (Fig.~\ref{fig:stopStart} (a)), the front of
the first vehicle crosses the stopping line about \unit[1.5]{s} after
the change to green corresponding to $\tau\approx\unit[0.7]{s}$ (the
rest of the time is needed to move the first meter to the
stopping line). If this vehicle started one second earlier, i.e.,
before the switching to green
(Fig.~\ref{fig:stopStart} (b)), the situation is yet save but an average
of 0.5 additional vehicles can pass during one green phase assuming
an outflow of \unit[1\,800]{veh/h} after some vehicles. Considering
the 12~vehicles that would pass in the reference scenario during the
\unit[30]{s} long green phase of the \unit[60]{s} cycle, this amounts,
on average, 
to an increase by \unit[4]{\%}.
An additional second can be
saved, allowing 13 instead of 12~vehicles per green phase,  if
the first vehicle stops \unit[4]{m} upstream of the stopping line
(instead of \unit[1]{m}) allowing an even earlier start without
compromising the safety
(Fig.~\ref{fig:stopStart}(d)). However, there are limits in terms of
acceptance and available space, so stopping \unit[2]{m} before the
stopping line (Fig.~\ref{fig:stopStart}(c)) is more realistic. In
effect, the 
latter strategy variants transform the anticipative start in a
``flying start'' which we will discuss now.

\subsection{\label{sec:flyingStart}Flying Start}

\begin{figure}[!ht]
  \begin{center}
    \includegraphics[width=0.6\textwidth]{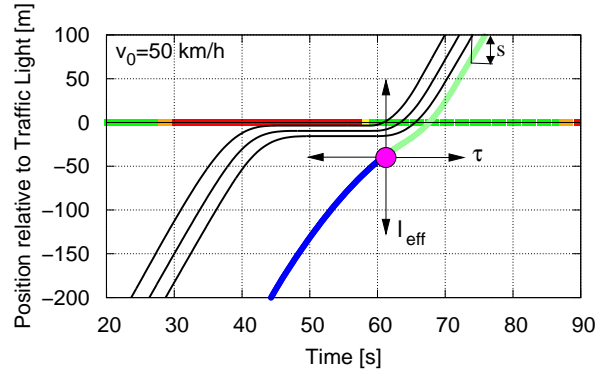}
  \end{center}
  \caption{Preemptive braking to avoid a stop: 
   Spatiotemporal target for the 4$\sup{th}$ vehicle (pink
   circle). The arrows indicate how the target changes when
   varying the reaction time $\tau$ or the
   effective length $l\sub{eff}$.}
  \label{fig:flyingStart}
\end{figure}

If, relative to the phases,  a vehicle arrives later than in the
previous two situations but too early to have a free passage,
preemptive braking may avoid a stop or, at least,
increase the minimum speed during the approaching phase. As depicted
in Fig.~\ref{fig:flyingStart}, the strategy
consists in controlling the vehicles's ACC such that a certain
spatiotemporal \emph{target point}  $(\Delta
  x,\Delta t$) relative to the stopping line and the switching time to
  green is reached. This point is determined such that a minimum of
  speed reduction is realized without impairing
  traffic efficiency by detaching this vehicle  from the platoon of
  leaders. 
 
From basic kinematic theory~\cite{Lighthill-W} and the properties of the IDM it
follows that the propagation 
velocity $c$ of the \emph{positions} of the vehicles at the respective
starting times is
constant and given
by $c \approx
-l\sub{eff}/\tilde{T}$ where $\tilde{T}$ is of the order of the IDM
parameter $T$. Assuming a gap $s_0^*$ of the first waiting vehicle to
the stopping line and a reaction delay $\tau$ of its driver, the
estimated spatiotemporal \emph{starting point} of the 
$n\sup{th}$ vehicle reads
\be
\label{targetPoint}
(\Delta x, \Delta t)=(s_0^*+(n-1)\, l\sub{eff}, \tau+(n-1) \, \tilde{T}).
\ee
The points lie on a straight line which is consistent with observations  (filled circles in
Fig.~\ref{fig:calibr}). While we assume that, by additional V2X communication from a
stationary detector to the vehicle, the equipped vehicle knows its order number $n$,
there are uncertainties in $\tau$, $l\sub{eff}$, and $T$ which depend on
unknown properties of the vehicles and drivers ahead. Furthermore, since the
strategy tries to avoid a stop, the  
\emph{target} point lies several meters upstream of and/or a few seconds
after the anticipated starting point. 

\begin{figure}[!ht]
  \begin{center}
    \includegraphics[width=0.9\textwidth]{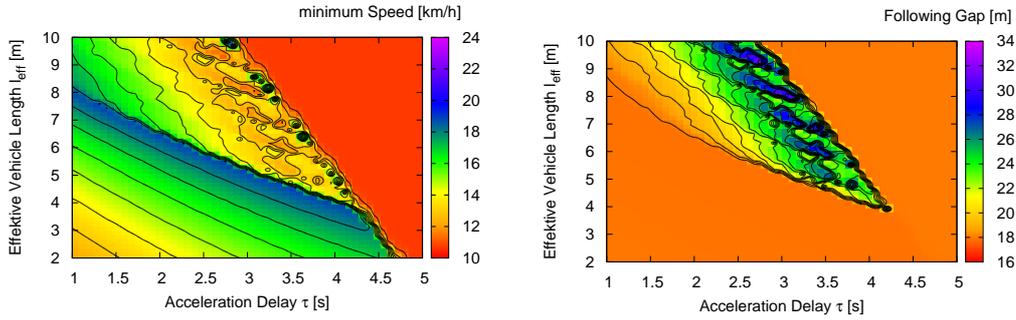}
  \end{center}
  \caption{Robustness of the premptive braking strategy. Shown is its
    efficiency for the $3\sup{rd}$ vehicle in terms of
    the minimum speed during the approach (left) and the gap
    once this vehicle is \unit[50]{m} downstream of the traffic light
    (right).} 
  \label{fig:flyingStartTauLeff}
\end{figure}

Is this strategy
nevertheless robust? In order to assess this, we treat $\tau$ and
$l\sub{eff}$ as free parameters of~\refkl{targetPoint} 
to be estimated and plot the performance metrics
minimum speed $v\sub{min}$ characterizing driving comfort and spatial gap $s$ to the platoon
(cf. Fig.~\ref{fig:flyingStart}) characterizing the dynamic capacity 
 as a function of $\tau$ and 
$l\sub{eff}$. 

Figure~\ref{fig:flyingStartTauLeff} shows these metrics
for the $n=3\sup{rd}$ vehicle arriving  at a timing such that
the minimum speed would be 
$v\sub{min}=\unit[10]{km/h}$ if this vehicle were not equipped. For the best
estimates (e.g., $l\sub{eff}=\unit[6.5]{m}$ and $\tau=\unit[2]{s}$), this minimum speed is nearly doubled without compromising
the capacity which would be indicated by an increased following
gap $s$. The simulations also show that estimation errors have one of three
consequences: (i) if the queue length and
dissolution time are estimated too optimistically ($l\sub{eff}$ and
$\tau$  too small), there is still a
positive effect since the minimum speed is increased without
jeopardizing the efficiency; (ii) if the queue is massively overestimated
($l\sub{eff}$ and $\tau$ significantly too large), the whole strategy is deemed unfeasible
and the approach reverts to that of non-equipped
vehicles; (iii) if, however, the queue is only slightly overestimated,
the strategy kicks in ($v\sub{min}$ increases) but the capacity is
reduced  since $s$ increases as well: the car does no longer catch the
platoon. A look at the parameter ranges (the plots range over factors of five in both
$\tau$ and $l\sub{eff}$) indicates that this strategy is
robust when erring on the optimistic side, if there is any doubt. 

Finally, we mention that counting errors (e.g. due to a vehicle changing
lanes when approaching a red traffic light meaning that this vehicle
has not passed the correct stationary detector) will lead to similar errors
for the estimated target point as above. Consequently, this strategy should be
robust with respect to counting errors as well.

\subsection{\label{sec:complex}Complex Simulation}

In the previous sctions, we have investigated the different strategies
of the TLA in isolation. However, there are interactions. For example, the
optimal target point of the flying-start strategy is shifted backwards
in time when equipped leading vehicles apply the
anticipative-start strategy. Furthermore, the question remains if the
TLA remains effective if there is significant surrounding traffic (up
to the level of saturation) and whether the results are sensitive to the
order in which slow and fast, equipped and non-equipped vehicles
arrive.

 We investigate this by complex simulations
of all strategies over several cycles where we vary, in each
simulation, the overall traffic 
demand (inflow) $Q\sub{in}$, and the penetration rate $p$ of equipped vehicles. Unlike
the simulations of single strategies, we allow for full stochasticity
in the vehicle composition. At inflow, we draw, for each new vehicle, the model parameters
from the independent uniform distributions specified in Sect.~\ref{sec:sim} and
assign, with a probability $p$, the property ``is
equipped''. 

\begin{figure}[!ht]
  \begin{center}
    \includegraphics[width=0.9\textwidth]{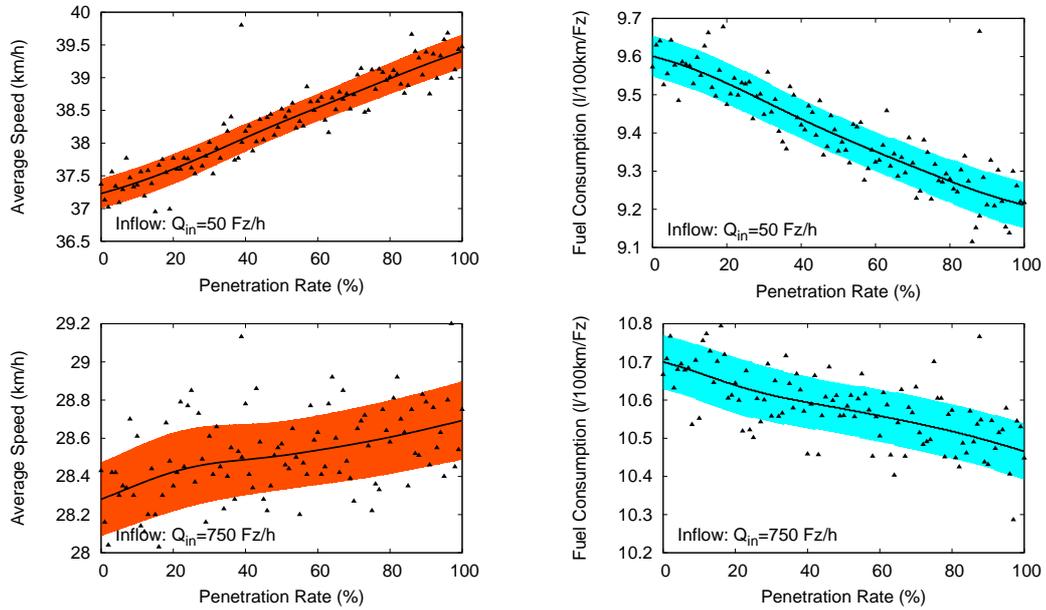}
  \end{center}
  \caption{Complex simulation of the overall effectiveness for all
    vehicles over several  cycles (see the main text for details).}
  \label{fig:complexRun}
\end{figure}

Figure~\ref{fig:complexRun} shows the results for the performance
metrics ``average speed'' (which is related to the average travel time),
and ``average consumption'' as a function of the penetration rate for
a small traffic demand (top row), and a demand near saturation
(bottom). Each symbol corresponds to a simulation for given values of
$Q\sub{in}$ and $p$. Due to the many stochastic factors and
interactions, we observe a wide scattering. Determining the local average
(solid lines) and $\pm 1 \sigma$
bands (colored areas) by kernel-based
regression (kernel width
\unit[15]{\%}), we nevertheless detect significant systematic
effects. For low traffic demand, we observe that both travel times and
fuel consumption are reduced by about \unit[4]{\%} when going from the
reference to $p=\unit[100]{\%}$
penetration. Furthermore, the effects essentially increase linearly with $p$, so the
\emph{relative performance indexes} $I\sub{T}$ and $I\sub{C}$ with respect to
travel time $T_t$ and fuel consumption $C$,
\be
\label{perfIndex}
I\sub{T_t}=-\frac{1}{T_t}\ablpart{T_t}{p}, \quad
I\sub{C}=-\frac{1}{C}\ablpart{C}{p}
\ee
are both constant and of the order of \unit[4]{\%}. Similar
performance indexes are obtained for the performance metrics ``number
of stops''. For higher traffic demand (lower row of
Fig.~\ref{fig:complexRun}), the relative performance of the 
TLA decreases except for the metrics ``dynamic capacity''.

\section{Discussion}

We have investigated, by means of simulation, a concept of a
traffic-light assistant (TLA) containing three strategies to optimize
the approach to and starting from traffic lights:
``economic approach'', ``anticipative start'', and ``flying
start''. The strategies are based on V2X communication: In order to
implement the TLA, equipped vehicles must obtain switching information
of the relevant traffic lights and -- as in the self-controlled signal
strategy~\cite{lammer2008SelfControl} -- counting data from a detector at
least \unit[100]{m} upstream of the traffic light. Complex simulations
including all interactions show that, for comparatively
low traffic demand, the TLA is effective. To quantify this, we
introduced relative performance indexes which we consider to be the most
universal approach
to assess penetration effects of individual-vehicle based ITS. For our specific setting
(maximum speed \unit[50]{km/h}, cycle time \unit[60]{s}, green time \unit[30]{s}), we obtained performance indexes of
about \unit[4]{\%} for most metrics if traffic demand is low. We
obtain higher values for higher maximum speeds and lower cycle times,
and lower values for a higher demand.  While the relative performance
is generally lower than that of the
traffic-adaptive ACC on freeways (about
\unit[25]{\%})~\cite{kesting-acc-roysoc}, the
\emph{individual} 
advantage kicks in with the first equipped vehicle, in contrast to
traffic-adaptive ACC.

\subsubsection*{Acknowledgements}
We would like to thank the Volkswagen AG who has sponsored part of
this work in a project.

\bibliographystyle{/home/staff/treiber/tex/inputs/alphadin}
\bibliography{MT-ITS-Treiber-refs}


\end{document}